\documentclass[8.5pt,twoside]{article}
\usepackage[super,sort&compress,comma]{natbib}
\usepackage{times,mathptm}
\usepackage{sectsty}
\usepackage{balance}
\usepackage[text={11.3cm,20.4cm},centering]{geometry}
\usepackage{graphicx} 
\usepackage{lastpage}
\usepackage[format=plain,justification=raggedright,singlelinecheck=false,font=small,labelfont=bf,labelsep=space]{caption}
\usepackage{subfigure}
\usepackage{graphicx,amscd,amsmath,amssymb,verbatim,amsfonts}
\usepackage{fancyhdr}
\let\vec=\mathbf

\begin{document}

\thispagestyle{plain}
\fancypagestyle{plain}{
\fancyhead[L]{\includegraphics[height=8pt]{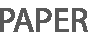}}
\fancyhead[C]{\hspace{-1cm}\includegraphics[height=15pt]{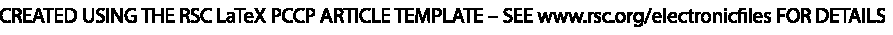}}
\fancyhead[R]{\includegraphics[height=10pt]{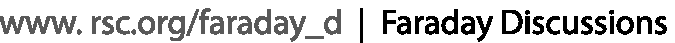}\vspace{-0.2cm}}
\renewcommand{\headrulewidth}{1pt}}
\renewcommand{\thefootnote}{\fnsymbol{footnote}}
\renewcommand\footnoterule{\vspace*{1pt}%
\hrule width 11.3cm height 0.4pt \vspace*{5pt}}
\setcounter{secnumdepth}{5}

\makeatletter
\renewcommand{\fnum@figure}{\textbf{Fig.~\thefigure~~}}
\def\subsubsection{\@startsection{subsubsection}{3}{10pt}{-1.25ex plus -1ex minus -.1ex}{0ex plus 0ex}{\normalsize\bf}}
\def\paragraph{\@startsection{paragraph}{4}{10pt}{-1.25ex plus -1ex minus -.1ex}{0ex plus 0ex}{\normalsize\textit}}
\renewcommand\@biblabel[1]{#1}
\renewcommand\@makefntext[1]%
{\noindent\makebox[0pt][r]{\@thefnmark\,}#1}
\makeatother
\sectionfont{\large}
\subsectionfont{\normalsize}

\fancyfoot{}
\fancyfoot[LO,RE]{\vspace{-7pt}\includegraphics[height=8pt]{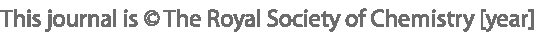}}
\fancyfoot[CO]{\vspace{-7pt}\hspace{5.9cm}\includegraphics[height=7pt]{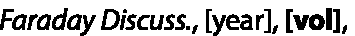}}
\fancyfoot[CE]{\vspace{-6.6pt}\hspace{-7.2cm}\includegraphics[height=7pt]{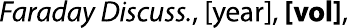}}
\fancyfoot[RO]{\scriptsize{\sffamily{1--\pageref{LastPage} ~\textbar  \hspace{2pt}\thepage}}}
\fancyfoot[LE]{\scriptsize{\sffamily{\thepage~\textbar\hspace{3.3cm} 1--\pageref{LastPage}}}}
\fancyhead{}
\renewcommand{\headrulewidth}{1pt}
\renewcommand{\footrulewidth}{1pt}
\setlength{\arrayrulewidth}{1pt}
\setlength{\columnsep}{6.5mm}
\setlength\bibsep{1pt}

\noindent\LARGE{\textbf{Electrostatic interaction of heterogeneously charged surfaces with semipermeable membranes}}
\vspace{0.6cm}

\noindent\large{\textbf{Salim R. Maduar,\textit{$^{a,b}$} Vladimir
Lobaskin,\textit{$^{c}$} and
Olga I. Vinogradova$^{\ast}$\textit{$^{a,b,d}$}}}\vspace{0.5cm}

\noindent\textit{\small{\textbf{Received Xth XXXXXXXXXX 20XX, Accepted Xth XXXXXXXXX 20XX\newline
First published on the web Xth XXXXXXXXXX 200X}}}

\noindent \textbf{\small{DOI: 10.1039/c000000x}}
\vspace{0.6cm}

\noindent \normalsize{In this paper we study the electrostatic interaction of a heterogeneously charged wall with a neutral semipermeable membrane. The wall consists of periodic stripes, where the charge density varies in one direction. The membrane is in a contact with a bulk reservoir of an electrolyte solution and separated from the wall by a thin film of salt-free liquid. One type of ions (small counterions) permeates into the gap and gives rise to a distance-dependent membrane potential, which translates into a repulsive electrostatic disjoining pressure due to an overlap of counterion clouds in the gap. To quantify it we use two complementary approaches. First, we propose a mean-field theory based on a linearized Poisson-Boltzmann equation and Fourier analysis. These calculations allow us to estimate the effect of a heterogeneous charge pattern at the wall on the induced heterogeneous membrane potential, and the value of the disjoining pressure as a function of the gap. Second, we perform Langevin dynamics simulations of the same system with explicit ions. The results of the two approaches are in good agreement with each other at low surface charge and small gap, but differ due to nonlinearity at the higher charge. These results demonstrate that a heterogeneity of the wall charge can lead to a huge reduction in the electrostatic repulsion, which could dramatically facilitate a self-assembly in complex synthetic and biological systems.
}
\vspace{0.5cm}

\section{Introduction}

\footnotetext{\textit{$^{a}$~A.N.~Frumkin Institute of Physical
Chemistry and Electrochemistry, Russian Academy of Sciences, 31
Leninsky Prospect, 119991 Moscow, Russia. }}
\footnotetext{\textit{$^{b}$~Faculty of Physics, M. V. Lomonosov Moscow State University, 119991 Moscow, Russia. }}
\footnotetext{\textit{$^c$~School of Physics and Complex and Adaptive Systems
Laboratory, University College Dublin, Belfield, Dublin 4,
Ireland}}
\footnotetext{\textit{$^{d}$~DWI,  RWTH Aachen, Forckenbeckstr. 50, 52056 Aachen, Germany}}

Long-range electrostatic interactions between surfaces play a central role in a variety of biological processes and substantially influence the properties of colloidal suspensions, thin films, and nanostructured materials. Most theoretical and experimental studies of electrostatic forces have been conducted for symmetric systems, and by assuming that the surfaces are homogeneously charged and impermeable.~\cite{Israelachvili2011} In this paper we focus on an  asymmetric case  of interactions of a patterned impermeable wall with a neutral semi-permeable membrane, bounding an aqueous electrolyte solution.

Surfaces with inhomogeneous charge distributions are of importance for several reasons. First, such inhomogeneous systems are ubiquitous, especially in biology. The best-known examples are the proteins, cellular membrane lipids,~\cite{Simons2000:lipid_rafts} soft anisotropic materials,~\cite{Glotzer2007} and self-assembled molecular layers on charged surfaces.~\cite{Meyer2005:exp_inhomo_force,borkovec:dendrimers_adsoption,LRattraction:Chen2006}
Second, the methods of surface treatment have advanced considerably during the past decade and enabled a fabrication of charge patterns in systems like as spherical Janus particles~\cite{Walther2008:Janus} and various patchy objects.~\cite{Du2011:anisotrop,Pawar2010} Third, they model the effect of defects in homogeneous systems.
In an effort to better understand the connection between a heterogeneity of a charge distribution and the amplitude of repulsive forces, the interaction of patterned walls in a liquid has been studied by several groups. Experimental studies have been performed to quantify the interaction between charged~\cite{Meyer2005:exp_inhomo_force,borkovec:dendrimers_adsoption,LRattraction:Chen2006} and neutral surfaces, where the average charge is zero,~\cite{Vreeker1992:stability_hetero,LRattraction:Perkin2006} with different distributions of surface charge heterogeneities. Most of theoretical effort has been focussed on the interaction between two periodically patterned surfaces~\cite{muller.vm:1983,Miklavic:hetero_94,Miklavcic_preturb} by using the linearized Poisson-Boltzmann (PB) equation, and boundary conditions of a fixed surface charge density (or a fixed surface potential) the heterogenous patterns. These studies concluded that for heterogeneously charged surfaces with a non-zero  total charge the leading-order interaction is dominated by the average charge, and that the repulsion between the surfaces becomes weaker than between two uniformly charged surfaces with the same average charge. However, for overall neutral surfaces the interaction of charge patches depends on the location and periodicity of the pattern and can change from being repulsive to an attractive one. Another recent development includes investigations of systems with randomly distributed charges,~\cite{Ben-Yaakov2013:hetero} strong correlations in systems with mobile charges, ~\cite{brewster2008:mobile_heterogeneous,pincus2011:heterogeneous_membranes} charge regulation, and non-linear ionic screening in heterogeneous systems.~\cite{Chg_regulation:van_Roij}

Donnan equilibria, which arise in the presence of semipermeable membranes, are of considerable importance in many areas of science and technology. Well known examples of semipermeable membranes are synthetic liposomes with ionic channels,~\cite{lindemann.m:2006} and multilayer shells of
polyelectrolyte microcapsules.~\cite{donath.e:1998,vinogradova.oi:2006,kim.bs:2007,vinogradova.oi:2004} Biological examples include viral capsids,~\cite{odijk.t:2003} cell~\cite{alberts.b:1983} and bacterial~\cite{sen.k:1988,stock.jb:1977,sukharev.s:2001} membranes. Since it was discovered, the theories of Donnan equilibria mainly focussed on the case of a single membrane,~\cite{zhou.y:1988,deserno.m:2002} or a single vesicle/capsule.~\cite{zhou.y:1988,tsekov.r:2006,stukan.mr:2006,tsekov.r:2008,siber.a:2012} However, ion equilibria play a very important role in processes involving interactions with a  membrane, such as adhesion to the wall. The quantitative understanding of electrostatic interactions involving membranes is still challenging. Previous investigations were restricted to interactions of two model membranes and relied on a number of assumptions and simplifications. Some solutions of the PB equation are known for charged bearing ionizable groups immersed in the salt reservoirs.~\cite{nimham.bw:1971} Later works in this direction assumed that the membranes are uncharged, and are separated by a thin film of salt-free solvent.~\cite{NLPB_sim2012} Results were not limited by calculations within the PB theory, and also included the Langevin dynamics simulations with explicit ions. In the wide gap limit, a repulsive disjoining pressure was predicted. Recent integral equation study suggested charge correlation effects in large concentration solutions of multivalent ions that could result in short-range attractions.~\cite{lobaskin.v:2012} We are unaware of any previous work that has addressed the question of interaction of a semipermeable membrane with a wall.

In this paper, we explore the charge and potential distributions arising when a neutral semipermeable membrane bounding the electrolyte solution, is separated by a thin of background solvent from the charged wall decorated by stripes with fixed densities of a local surface charge. We first solve analytically a linearized PB equation for a weak local charge, and evaluate the distribution of electrostatic potential in the system. We then derive an explicit expression for a pressure on the membranes and a disjoining pressure in the gap between them. Our mean-field approach
is verified by Langevin dynamics simulations.

\section{Theory}

\begin{figure}[h]
\begin{center}
\includegraphics [width=7 cm]{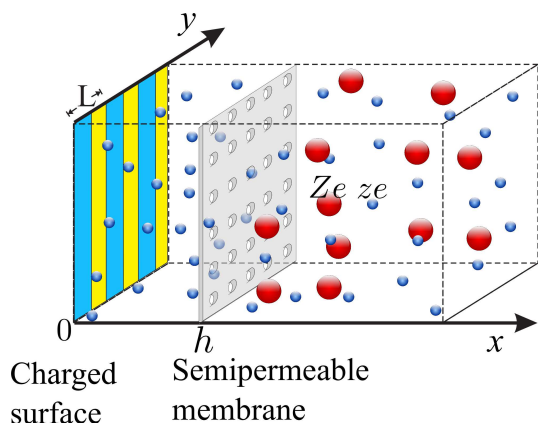}
   \end{center}
  \caption{Schematic of the studied system consisting of neutral semipermeable membrane at $x=h$ with a heterogeneously charged surface  at $x=0$.
   The period of the charge distribution is denoted by $L$. Small spheres indicate small ions. The
large ions are also depicted as spherical, which is appropriate, for instance, for conventional charged colloids, nanogels or micelles, but our conclusions are general. They
could also apply for cylindrical
e.g., DNA, viruses, actin filaments
or
polyelectrolytes.}
  \label{fig:problem}
\end{figure}

We consider a system consisting of a charged impermeable wall and a semipermeable membrane in a contact with an electrolyte solution. The gap between the wall and the membrane, $h$,  is filled with a salt-free solvent (Fig.~\ref{fig:problem}). We assume that the membrane is permeable for one type of ions (small ions or counter-ions) with charge $z e$ and impermeable for another type (large ions) with charge $|Z e| \ge |z e|$. Here, $Z$ and $z$ are valencies of large and small ions, respectively, and $e$ is the elementary charge. The ion concentrations are denoted by $C$ for large ions and $c$ for small ions. The membrane is infinitesimally thin, rigid, and electrically neutral.
We focus on periodic, charged, striped wall with an average charge density $\sigma^s$, where the charge and the potential, $\psi$, are varying in only one direction, $y$, with a periodicity $L$. Alternating stripes are characterized by charge densities $\sigma^s_1$ and $\sigma^s_2$. The surface fraction of stripes of type 1 is denoted as $\omega = L_1/L$, where $L_1$ is the width of the stripe with charge density $\sigma^s_1$. The permittivities of inner and outer solutions are equal and denoted below as $\varepsilon$.

\subsection{Potential}

We first introduce the dimensionless electrostatic potentials $\varphi_{i,o}={z e\psi_{i,o}\over k_BT}\ll 1$ with the index ${i, o}$ standing for ``in'' ($ x < h$) and ``out'' ($ x \ge h$) of the confined slab.~\cite{NLPB_sim2012}
We then assume a weakly charged surface, so that $\varphi$ satisfies the linearized PB approach:
\begin{eqnarray}
&\Delta\varphi_i (x,y)&={\kappa_{i}}^2 (\varphi_i(x,y)-1), \quad 0<x<h \label{DH1}\\
&\Delta\varphi_o (x,y)&={\kappa_{o}}^2\varphi_o(x,y), \qquad \qquad \quad x \ge h, \label{DH2}
\end{eqnarray}
where the inner inverse  screening length, $ \kappa_i^{-1}$, is defined as $\kappa_i^2= 4\pi \ell_B c_0$ with $\ell_B=z^2e^2/(4\pi\varepsilon k_BT)$ the Bjerrum length, $\tilde Z=Z/z <0$ is the valence ratio of large and small ions, and $ c_0$ is the bulk concentration of small ions in the outer space. The outer inverse screening length, $\kappa_o$, can be calculated as $\kappa_{o}^2 = 4\pi \ell_B (\tilde Z^2 C_{0}+c_0)$, where $ C_0$ is the concentration of large ions far from the membrane. Obviously, it represents the inverse Debye length of the bulk electrolyte solution in the container. Enforcing the electroneutrality $Z C_0 + z c_0 = 0$, we find $\kappa_o = \kappa_{i} \eta$, where $\eta=\sqrt{1-\tilde{Z}}$. Note that for this particular problem, the main reference length scale that determines the behavior of the system is $\kappa_{i}^{-1}$, as we show below.

We solve these equations with a boundary condition of prescribed surface charge $\sigma^s(y)$ on the wall and continuity of the electric field at $x=h$, which corresponds to the case of a neutral membrane:
\begin{eqnarray}
&\partial_x\varphi_i(x,y)\vert_{x=0}& = - \frac{4\pi ze}{\varepsilon k_BT}\sigma^s(y) = - b(y),
\\
&\partial_x\varphi_i(x,y)\vert_{x=h}& = \partial_x\varphi_o(x,y)\vert_{x=h},
\label{BC}
\end{eqnarray}
where $b(y) = \frac{4\pi ze}{\varepsilon k_BT}\sigma^s(y)$ is a local analogue of the Gouy-Chapman inverse length. In our case of alternating stripes $b(y)$ switches between two values, $b_1$ and $b_2$. Parameter $ b_{1,2} \kappa_i^{-1}$ then characterizes the interplay between ion-ion and ion-wall interactions.~\cite{Nato_andelman}

Note that at high charge densities and high values of the electric potential, the description of the problem cannot be simplified
by linearization of the PB approach. Beside that, correlations between macroions should be taken into account in the limit of large charges $Z$. Based on earlier results,~\cite{tsekov.r:2008} one can expect to observe the same qualitative picture at least at low polyion concentrations, while at the higher concentrations the correlation effects might become significant.~\cite{lobaskin.v:2012} We leave the study of the latter regime for a future work.

Applying
boundary conditions \eqref{BC} to Eqs. (\ref{DH1}), (\ref{DH2}) we obtain a distribution of the potential:
\begin{equation}
\begin{split}
\varphi_i(x,y)&=1+\frac{b_0 \kappa_i^{-1}\cosh[\kappa_i(h-x)]-\eta_0 \cosh[\kappa_i x]+\eta_0 b_0 \kappa_i^{-1}\sinh[\kappa_i(h-x)]}
{\eta_0\cosh[\kappa_i h]+\sinh[\kappa_i h]}\\&
+\sum\limits_{n\neq 0}\frac{b_n}{q_n} \frac{\cosh[q_n(h-x)]+
\eta_n \sinh[q_n(h-x)]}{\eta_n\cosh[q_n h]+\sinh[q_n h]} e^{i k_n y},\label{phiCC}
\end{split}
\end{equation}
where  $b_n$ is the Fourier coefficient of $b(y)$, $k_n=\dfrac{2\pi n}{L}$, $\ q_n^2 =k_n^2+\kappa_i^2$, and $\eta_n^2=\dfrac{k_n^2+\kappa_o^2}{q_n^2}$. The average dimensionless surface charge is $b_0 \kappa_i^{-1}$, where $b_0=b_1 \omega + b_2 (1 - \omega)$.

\subsection{Disjoining pressure}

At the equilibrium, the disjoining pressure consists of two parts, namely, the pressure due to electric volume force ($\rho \vec{E}$) and the ideal osmotic pressure.~\cite{Ohshima2010:plane_out} Within the linearized PB theory, one should replace the boundary density rule from the nonlinear PB theory by its linear case analogue.~\cite{tsekov.r:2008,Deserno2002:LPB_unified_approach} Below we discuss this for our system.

A mechanical equilibrium requires that the solution for the potential and charge distribution satisfy the hydrostatic equation
\begin{equation}
0 = -\vec{\nabla} p + \rho \vec{E} = \vec \nabla \cdot \left(\mathbb{T}- \mathbb{I} \ p\right) \equiv- \vec \nabla \cdot \Pi \label{hydrostatic}
\end{equation}
where $\mathbb{T}$ is the Maxwell's electrostatic stress tensor
\begin{equation}
\mathbb{T}_{ij} = \dfrac{\varepsilon}{4\pi}\left[E_i E_j-\delta_{ij} \dfrac{\vec{E}^2}{2}\right]
\end{equation}

The difference, $\Pi(x,y) = \mathbb{T}(x,y)-\mathbb{I} \ p(x,y)$, represents an electrostatic disjoining pressure, which is equal to the excess osmotic pressure at the particular position in the film, $x_0$, where the magnitude of potential has a minimum value, and electrostatic stress vanishes ($\mathbb{T}=0$).~\cite{Andelman:2010_PB_revisit,Andelman_biophysics,Nato_andelman,NLPB_sim2012} We remind that for heterogeneous objects is a surface integral of $\Pi\cdot \vec n_s$ with respect to surface normal $\vec n_s$.~\cite{Russel1989,SEI97,stankovich1996} Therefore, to find $x_0$ for our system, which now could depend on $y$, we propose to use $y$-average disjoining pressure $\dfrac{1}{L}\int\limits_{y=0}^{y=L}\Pi (x,y)\equiv \left<\Pi (x,y)\right>_y$, as a measure of an electrostatic interaction.
We then use Eq.~\eqref{hydrostatic} to calculate the average disjoining pressure
\begin{equation}
\left<\Pi (x_0,y)\right>_y = \left<p - \mathbb{T}\right>_y =\left< p(x_0,y)\right>_y \approx k_BT c_0\left(1-\left<\varphi_i(x_0,y)\right>_y\right) \label{Eq:disj_zerofield}
\end{equation}

To calculate the value of a potential at the (still unknown) point,  $x_0,y_0$, we have to find a relation between the field and the potential. This can be done by multiplying  Eq.\eqref{DH1} by $\vec \nabla \varphi\equiv \{\partial_x \varphi, \partial_y \varphi\}$.  Taking into account that $\vec E=-\vec \nabla \varphi \dfrac{k_BT}{z e}$, we find:
 \begin{equation}
-\vec\nabla \left[ \dfrac{\mathbb{T}}{k_BTc_0} -   \left( \dfrac{\varphi_i^2(x,y)}{2} - \varphi_i(x,y) \right) \mathbb{I}\right]=0
\label{invariant_DH}
 \end{equation}
The off-diagonal components represent the \emph{tangential force} on charged surface, which should vanish on average. The $y$-averaging of Eq.~\eqref{invariant_DH} then eliminates these components of Maxwell stress tensor  $\left<\mathbb{T}_{x,y}\right>=
\left<\mathbb{T}_{y,x}\right>=0$:

\begin{equation}
\dfrac{\partial}{\partial x}\left< \dfrac{(\partial_x \varphi_i)^2- (\partial_y \varphi_i)^2}{2 \kappa_i^2} +  \left( \dfrac{\varphi_i^2(x,y)}{2} - \varphi_i(x,y) \right)\right>_y =0 \equiv \dfrac{\partial}{\partial x} C(h)
\label{hydrostatic_constant}
 \end{equation}
where $C(h)$ is an integration constant for the linearized PB equation. Comparing Eqs. (\ref{invariant_DH},\ref{hydrostatic_constant}) to Eq. \eqref{hydrostatic}, we derive
\begin{equation}
p(x,y) = A + k_BTc_0\left( \dfrac{\varphi_i^2(x,y)}{2} - \varphi_i(x,y) \right)
\end{equation}
Following the approach~\cite{tsekov.r:2008,NLPB_sim2012,Deserno2002:LPB_unified_approach} we find the constant $A$ by using the van't Hoff's law at the position of minimum-magnitude potential, $x_0$:
\begin{equation}
p(x,y) = k_BTc(x_0,y) +k_BTc_0\left( \dfrac{\varphi_i^2(x,y)-\varphi_i^2(x_0,y)}{2} - (\varphi_i(x,y)-\varphi_i(x_0,y)) \right)
\end{equation}

After solving the linearized PB equation for $ \varphi$, one can verify that the left-hand side of Eq.\eqref{hydrostatic_constant}, \textit{i.e.} $C(h)$, does not depend on $x$ but can be a function of separation, $h$. Without loss of generality, we then can calculate the constant by using Eq.\eqref{hydrostatic_constant} at $x=h$:

\begin{equation}
C(h) = \left<\dfrac{\tilde{Z}}{2} {{\varphi}^m(y)}^2 -{\varphi}^m(y)\right>_{y-average}, \label{Eq:constant}
\end{equation}
where ${\varphi}^m$ is a potential of a membrane.

Now we set $E_x=E_y=0$ in Eq.~\eqref{hydrostatic_constant} and solve it with respect to $\varphi(x_0)\equiv\varphi|_{\mathbb{T}(x_0,y)=0}$:
\begin{equation}
\left<\dfrac{1}2\varphi^2(x_0,y)-\varphi(x_0,y)\right>_y=C(h)
\end{equation}
By substituting this into Eq.\eqref{Eq:disj_zerofield}, we obtain an expression for disjoining pressure in the gap between the semipermeable membrane and the heterogeneously charged surface:
\begin{equation}
\left(\dfrac{\Pi(h)}{k_BTc_0}\right)^2 = 1+2C(h)=1+\left<{\tilde{Z}} {{\varphi}^m(y)}^2 -2 {\varphi}^m(y)\right>_{y-average}\label{Eq:disj_pressure}
\end{equation}
A remarkable corollary of this relation is that the disjoining pressure can be easily determined once the induced membrane potential is found.

\section{Simulation}

The Langevin dynamics (MD) simulations are performed on the level of the primitive model with explicit large and small ions using
the ESPResSo simulation package.~\cite{Espresso} Our model also includes charged surface and neutral semipermeable membrane.
The membranes are made impermeable for cations, but ``invisible'' for anions.~\cite{stukan.mr:2006}

All electrolyte ions repel each other with a repulsive Weeks-Chandler-Andersen (WCA) potential~\cite{weeks:5237} of range $\sigma_{WCA}$ and magnitude $\epsilon_{WCA}$. The same potential acts between particles and the walls (charged surfaces and semipermeable membranes). To illustrate our approach, we here use only a
monovalent electrolyte ($Z=1$, $z=-1$). The temperature is set by a Langevin thermostat to $k_BT=1.0 \epsilon_{WCA}$.

The solvent is treated as a homogeneous medium with a dielectric permittivity set through the Bjerrum length $\ell_B$. The electrostatic
interaction between the ionic species is modelled by the Coulomb potential $U_{\rm Coul}(r_{ij})=k_BT\frac{\ell_B q_i q_j}{r_{ij}},$
where $q_i=\pm1$ with $\ell_B =0.8\sigma_{WCA}$ to $1\sigma_{WCA}$. We model the systems with 2D-periodicity in $y$ and $z$ directions to exclude any boundary effects. The electrostatics is calculated using P3M~\cite{P3M1989} method combined with the electrostatic layer correction (ELC)-algorithm~\cite{ELC2002} with gap size $50\sigma_{WCA}$.

 Bulk ion concentrations vary  from  $10^{-4}\sigma_{WCA}^{-3}$ to $10^{-3} \sigma_{WCA}^{-3}$, which gives the screening length in the range $ \kappa_i^{-1} = 6\sigma_{WCA}$ to  $20\sigma_{WCA}$. We verified that the force exerted on the surface depends on the dimensionless parameter $\kappa_i h$ rather than on $\kappa_i$ or $h$ separately. Therefore for force measurements upon $\kappa h$ we fixed $ \kappa_i^{-1} = 10 \sigma_{WCA}$ and varied $h$ in the range from $3\sigma_{WCA}$ to $70\sigma_{WCA}$.  These values allows us to calculate the dependence of the interaction force for  a wide range of dimensionless separations $\kappa_i h=0.3-20$.

\emph{Charged plate} is constructed  from discrete charges at surface density $\sigma^s=10^{-2} ~q_s e\times\sigma_{WCA}^{-2}$, where $q_s e$ is the charge of a discrete surface ion. The surface charges are located at $x=0$ and random $\{y,z\}$ coordinates. In our system, the inverse Gouy-Chapman length is equal to $b=0.1 \sigma$ so that $b\kappa^{-1}=1$. To model heterogeneity we used periodic charge pattern  with stripe widths $L_1=L_2= 50\sigma$. This gives fraction of charged stripe $\omega=0.5$ at which heterogeneity effects are the most pronounced. Dimensionless periodicity could be varied in a wide range up to $\kappa_i L \approx 10$.

We use a \emph{simulation box} of depth $L_x=100\sigma_{WCA}-200\sigma_{WCA}$ in the $x$ direction, which is confined by impermeable walls at both ends ($x=0$ and $x=L_x$). The lateral dimensions $L_y \times L_z = 200\sigma_{WCA} \times 100\sigma_{WCA}$ and number of ions ($N = 1000-4500$) are selected large enough to achieve constant bulk ion concentrations at large $x$, far from the membrane. The width of the simulation box in the $y$-direction was set so that we had at least two periods of the charge pattern within the unit cell.

We evaluated the pressure on a plate by summation of contributions of all ions. For example, the Lennard-Jones interaction force excreted by an ion on the surface located at $x=h$ is equal to:
\begin{equation}\label{fx}
F(x) = 4 \epsilon_{WCA} \left(\frac{12\sigma_{WCA}^{12}}{\left(x-{h}\right)^{13}}-\frac{6\sigma_{WCA}^{6}}
{\left(x-{h}\right)^{7}}\right)
\end{equation}
The surface-averaged pressure
\begin{equation}
p=\int_{h/2}^{h+2^{1/6} \sigma} dx <C(x,y)>_y F(x).
\end{equation}

We measure both bulk osmotic pressure -- the pressure at the end of the box $x=L_x$ -- and the force on the membrane exerted by large ions.
The calculated bulk osmotic pressure is further used to derive bulk concentrations $c_0, C_0$ and to normalize disjoining pressure by factor $k_BTc_0$.

\section{Results and Discussion}

In this section we present some example calculations based on the analytical linearized PB theory and results of Langevin dynamics simulations.

\subsection{Homogeneously charged wall}

\begin{figure}[h]
\center{\includegraphics[width=10cm]{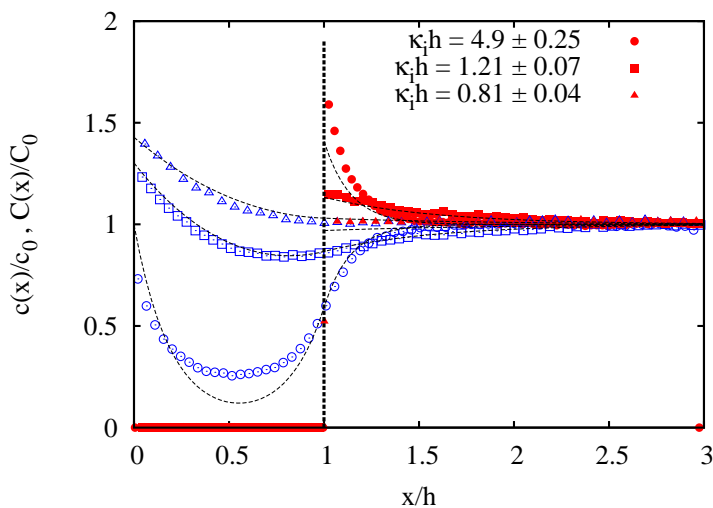}}
\caption{Distribution of small and large ions in the system. Dashed curves show predictions of the linearized PB theory. Symbols show simulation results. Open symbols indicate small ions, filled symbols correspond to large ions. }
\label{fig:homo:concentrations_simulation}
\end{figure}

We begin by studying the case of a homogeneously charged wall, which will be a reference system for our problem. Let us first focus on ion distributions in the system, which are shown in Fig.~\ref{fig:homo:concentrations_simulation} versus $x/h$
(symbols). Also included are the theoretical curve (solid curves).  The agreement is excellent at small $\kappa_i h$ (strong overlap of an inner double layer). Such a situation would be realistic for dilute solutions and/or thin gap.
At large $\kappa_i h$ (weak overlap of the inner ionic layers), i.e. thicker films and/or more concentrated solutions a linearized PB theory fails to describe quantitatevely the simulation data in the gap. In this case, the large ions are concentrated near the membrane, which is reflected by a very sharp concentration. The value of this peak calculated within linearized approach differ from the simulation value. Nevertheless, a linear theory is in a good qualitative agreement with simulation data, and could safely be used as a first approximation. We should like to stress that this is neither adsorption
driven by an attraction of ions to the membrane nor condensation driven by an attraction between ions. In our case, we deal with another
effect, where electrostatic self-assembly of large ions and a neutral membrane is caused by attraction of
large ions to inner counterions. This, in turn, is the consequence of a counterion leakage leading to an excess charge
of inner and outer regions.
At large $\kappa_i h$ concentration profiles of small ions have a minimum, which can be used to calculate a disjoining pressure. In case of simulation results we can employ a boundary density rule \begin{equation}
\Pi_0 = k_BTc_{min} = k_BTc_0+k_BTC_0 - k_BTC^m
\end{equation}
Here and below subscript $0$ of $\Pi$ corresponds to the case of a homogeneous wall.
Simulation data presented in Fig.~\ref{fig:homo:concentrations_simulation} are indeed in excellent agreement with this formula, but the linearized PB theory obviously deviates from its prediction (see~\cite{tsekov.r:2008,Deserno2002:LPB_unified_approach} for a detailed discussion of calculations of a pressure in the linearized PB theory).

\begin{figure}[h]
\begin{center}
\includegraphics[width=8cm,trim=1cm 0 0 0]{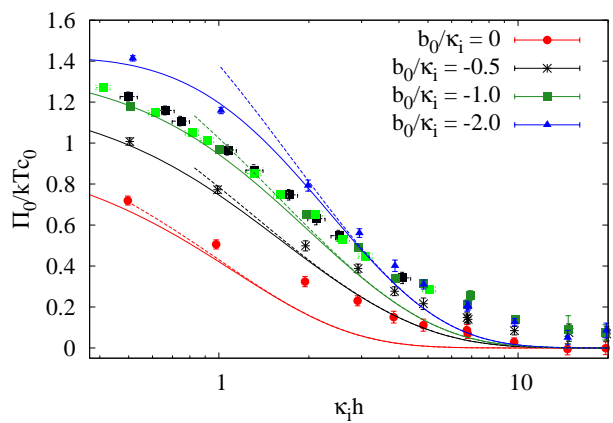}
\end{center}
\caption{Disjoining pressure in the gap between a homogeneously charged wall and a semipermeable membrane simulated at different surface charge densities on the wall (symbols). For a charge density of $  b_0 \kappa_i^{-1} =-1$ data were obtained at several $c_0$. Solid curves show predictions of linearized PB theory [Eq. \eqref{Eq:disj_pressure}]. Dashed curves are asymptotic results calculated with Eq.\eqref{homoCC_largesep}.  }\label{fig:homo:disjoining}
\end{figure}

Fig.~\ref{fig:homo:disjoining} shows simulation data for a disjoining pressure as a function of $\kappa_i h$ obtained for walls with different surface charge. Also included are theoretical results calculated with Eq. \eqref{Eq:disj_pressure}. The agreement between theory and simulations is quite good, but one can see that at large $\kappa_i h$ linear theory underestimates the value of the disjoining pressure. The data presented in Fig.~\ref{fig:homo:disjoining} show larger $\Pi_0$ at larger value of the surface charge $b_0 \kappa_i^{-1}$. Note that in case of uncharged wall, $b_0=0$, our system is equivalent to a symmetric system of two semipermeable membranes~\cite{NLPB_sim2012} separated by a twice larger distance, $2h$.

Finally, we note that simple asymptotic expressions can be constructed for large and small $\kappa_i h$. Thus, in the limits of large $\kappa_i h$ we derive
\begin{equation}
\left(\frac{\Pi_0(h)}{k_BT c_0}\right)^2\approx 4\dfrac{\eta^2}{(1+\eta)^2} e^{-2\kappa_i h}
-4\frac{b_0}{\kappa_i}\frac{\eta}{1+\eta} e^{-\kappa_i h} \label{homoCC_largesep}
\end{equation}
These asymptotic curves are included in Fig.~\ref{fig:homo:disjoining}. Eq.(\ref{homoCC_largesep}) indicates qualitatively different behavior of $\Pi_0(h)$ in case of neutral and charged walls. For a neutral wall the second term vanishes, and only the first term determines a decay of $\Pi_0(h)$. For charged walls the first term can safely be ignored, and asymptotics is determined by the second term. In the limit of small $\kappa_i h$ we obtain
\begin{equation}
\left(\frac{\Pi_0(h)}{k_B T c_0}\right)^2\approx 1 -
\frac{2\kappa_i b_0 \eta-b_0^2 \tilde{Z}}{\kappa_i^2\eta^2},\label{homoCC_smallsep}
\end{equation}
which gives the maximum value of a disjoining pressure in our system.

\subsection{Heterogeneously charged wall}

\begin{figure}

\begin{center}
\includegraphics[scale=0.7,trim=0cm 1.5cm 0cm 1.5cm]{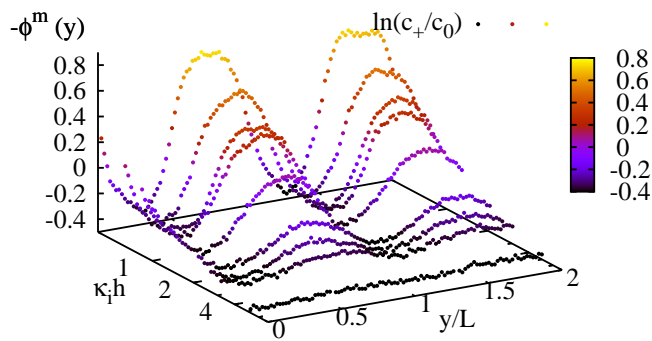}
\end{center}
\caption{Membrane potential calculated from concentration profiles collected from $x=h$ to $x=h+2.5\sigma$. The charge density of the surface stripes is given by $  b_{1,2} \kappa_i^{-1} = -1 \pm 2$.}\label{Fig:simulation_membrane_pot}
\end{figure}

\begin{figure}
    \centering
       \subfigure[$ b_{1,2} \kappa_i^{-1}= -1 \pm 0.5$]
    {
        \includegraphics[width=2in,trim=1cm 0.5cm 0cm 0cm]{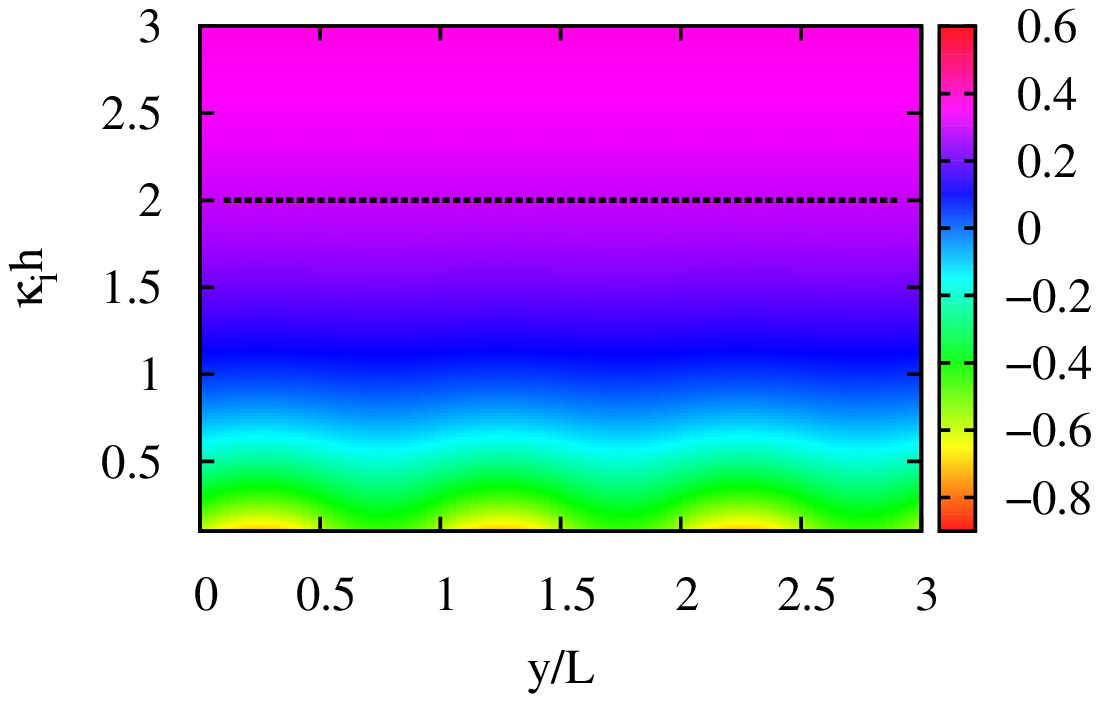}
        \label{fig:second_sub}    }
         \subfigure[$ b_{1,2} \kappa_i^{-1}= -1 \pm 2.5$]
    {
        \includegraphics[width=2in,trim=1cm 0.5cm 0cm 0cm]{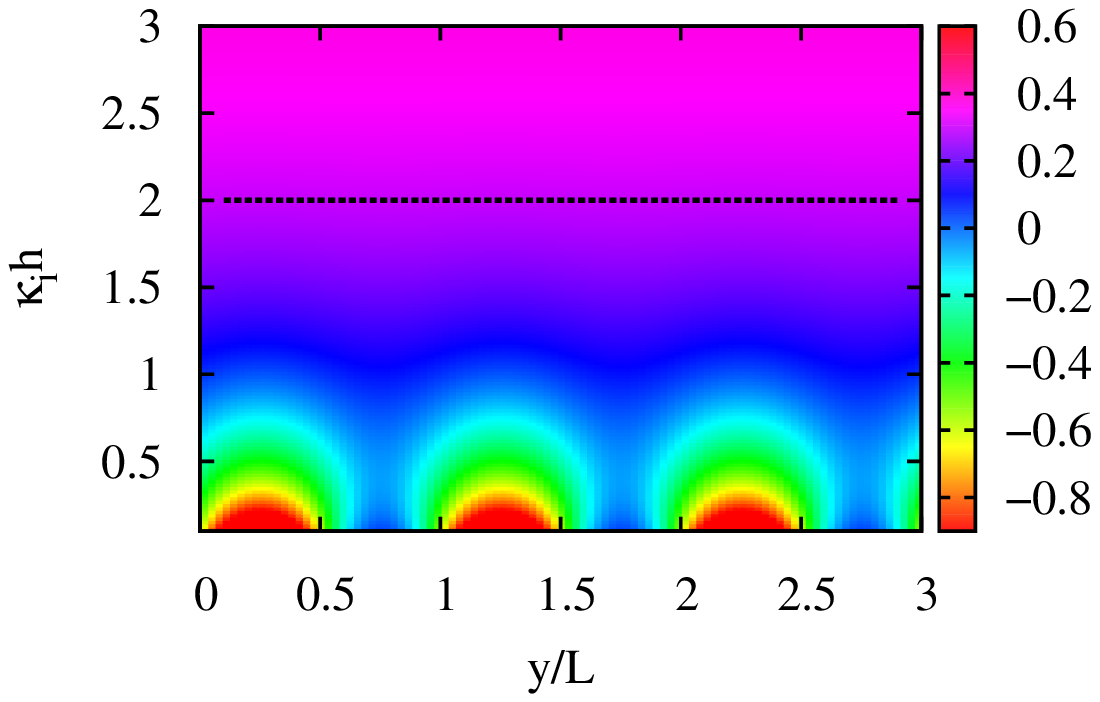}
        \label{fig:forth_sub}
    }
    \caption{Membrane potential, $\varphi^m$, (colorbar shows the color scale) as a function of $\kappa_i h$ and $y/L$ calculated at $\tilde{Z}=-1$, $\kappa_i L = 2$, and $  b_0 \kappa_i^{-1} = -1$. Dashed line shows the distance from the wall $h =L$.}\label{fig:inhomo:surf_pot}
\end{figure}

We begin by studying the membrane potential. Fig.~\ref{Fig:simulation_membrane_pot} plots results evaluated from simulated concentration profiles [as $\varphi^m=-\log(c/c_0)$]  collected in the interval of $\Delta x$ from $h$ to $h+0.25 \kappa_i^{-1}$. The simulation data demonstrate that a heterogeneously charged surface induces an inhomogeneous potential of the uncharged membrane if separations are small enough. Theoretical predictions shown in Fig.~\ref{fig:inhomo:surf_pot} are in good qualitative agreement with simulation results.

 Our results show that at $ h > L$ the membrane potential is uniform and does not vary with $y$, which is in agreement with earlier predictions made for impermeable surfaces.~\cite{Miklavic:hetero_94} At small $h<L$ there are pronounced variations of the induced membrane potential in the $y$-direction, and its sign coincides with that of the charge patches of the wall. Note that for a very small $\kappa_i h \ll 1$ and and strong screening, $\kappa_i L >1$,  the distribution of the membrane potential becomes locally uniform within each stripe $L_1$ or $L_2$. The net membrane potential is then given merely by a sum of independent contributions of each charged stripe. Such superposition approximation was previously used to describe heterogeneously charged impermeable walls separated by a film of an arbitrary thickness.~\cite{Kuin1990:superposition} We see that in case of a membrane the range of applicability of this model is much smaller.

\begin{figure}[h]
\begin{center}
\includegraphics[width=10cm]{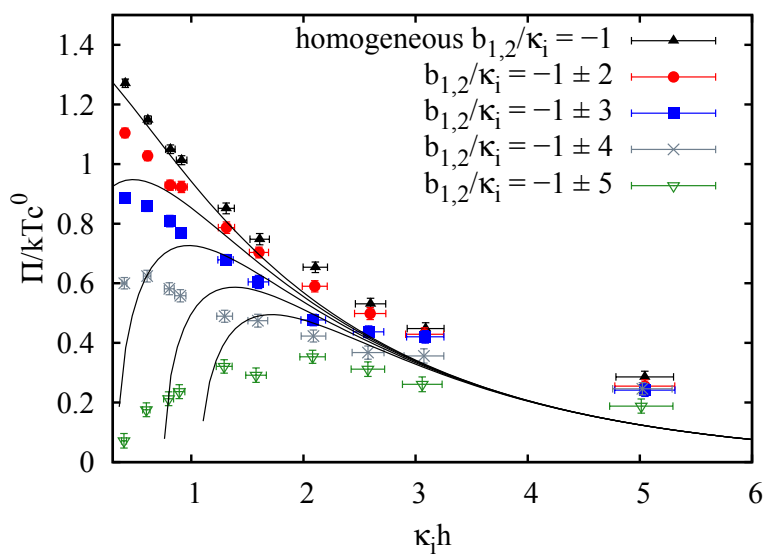}
\end{center}
\caption{Disjoining pressure in the gap between a heterogeneously charged wall and a semipermeable membrane ($\kappa_i L = 10$). Dashed curves show predictions of the linearized PB theory, symbols show the simulation results. }\label{fig:inhomo:forces}
\end{figure}

Fig.~\ref{fig:inhomo:forces} shows  the disjoining pressure in the gap between the striped wall and the membrane at fixed $\kappa_i L = 10$. We also fix an average charge of the wall, $b_0 \kappa_i^{-1}=-1$, and $\omega = 0.5$,  but vary the surface heterogeneity, $b_1 - b_2$. It can be seen that for large $\kappa_i h$ (above 3 for our parameters) the heterogeneity of the wall does not play any role. However, at smaller $\kappa_i h$ there is a discrepancy from the reference homogeneous system, especially where the heterogeneity is higher. The discrepancy is always in the direction of the smaller pressure than predicted for a homogeneously charged wall. Therefore, the contribution from the heterogeneity can be seen as an additional attractive force acting in the system. Simple arguments given below justify this conclusion. Indeed, if $\kappa_i h \gg 1$, one can derive
\begin{equation}
\left(\frac{\Pi(h)}{k_BT c_0}\right)^2\approx \left(\dfrac{\Pi_0(h)}{k_BTc_0}\right)^2 -
\frac{(b_1-b_2)^2}{\kappa_i^2} \frac{(\eta-1)}{\eta+1} \omega (1 -\omega) e^{-2 q_1 h}\label{heteroCC_kappa_largesep}
\end{equation}
 The second (negative) term can be interpreted as an exponentially decaying (weak) attractive force with the characteristic length $q_1^{-1}$. For $\kappa_i h \ll 1$, we get
\begin{equation}
\left(\frac{\Pi(h)}{k_BT c_0}\right)^2\approx \left(\dfrac{\Pi_0(h)}{k_BTc_0}\right)^2+ \frac{(b_1-b_2)^2}{\kappa_i^2} \frac{\tilde{Z}}{\eta^2} \frac{\omega ( 1- \omega)}{2}\label{heteroCC_kappa_smallsep}
\end{equation}
The second term of this expression is again negative (i.e. attractive), but of much larger amplitude, which allows to interpret the results presented in Fig.~\ref{fig:inhomo:forces}. In particular, it explains a stronger attraction for more heterogeneous (i.e. with larger $b_1 - b_2$) surfaces.

\section{Conclusions and Perspectives}

In this paper, we have considered the effect of surface charge heterogeneity on the electrostatic interaction with a neutral semipermeable membrane in a contact with a bulk reservoir of an electrolyte solution. Two approaches have been followed. First, we have used continuum electrostatics, namely, linearized PB approach,  to
propose a macroscopic estimate of the electrostatic disjoining pressure associated with a surface characterized by a heterogeneous (striped) charge pattern. This (analytical) approach has enabled us to determine the important factors
controlling the electrostatic interaction with a membrane. In particular, we have demonstrated that the membrane potential can be tuned by a charged wall located near a membrane, so that a membrane can take a heterogeneous electrostatic potential. We have also shown that surface heterogeneity becomes important at low net surface charge, large $\kappa_i L$ and relatively small, compared to $L$ distances. In this case a heterogeneity reduces repulsive disjoining pressure as compared with expected for a uniformly charged wall. In other situations the patterned surface can be treated as a homogeneous one. Then on the basis of Langevin dynamics simulations we have verified our theory for weakly charged surfaces and small  $\kappa_i h$. However, a discrepancy between the two approaches is exhibited at large $\kappa_i h$ and/or strongly charged surfaces. This points out the importance of the nonlinear effects in the full PB theory.

In our paper, we presented results on the repulsive interaction associated with a surface formed by alternating stripes
of different local charge and a neutral semi-permeable membrane in a contact with a simple water-electrolyte solution. However, our results can be easily extended to more complex patterns, relevant to experimental and biological systems, to charged membranes, and polyelectrolyte systems. As some examples, the heterogeneity may have a dramatic implication on the cell adhesion since it is known to be controlled by non-specific forces such as long-range electrostatics.~\cite{Kalasin2010:bacteri_hetero} Beside that, similar  to~\cite{Gingell1967:surface_potential_cell_interaction} we propose that  the neighbouring charged objects affect the state of cell membrane surface, cell interactions as well as complicated biological processes as endocytosis and signalling processes through altering ``cell electrostatic indicator'' -- membrane potential.
Finally, the induced potential could  change a conformation of membrane proteins, which in turn could affect the ion channels. Note that such complex systems cannot be solved in an analytical way. Still we believe that our study suggested a promising way towards understanding some basic physics underlying the behavior of these biological objects.

\section*{Acknowledgements}
This work was supported by the Russian Foundation for Basic Research (grant 12-03-00916) and SFB 985 ``Functional microgels and microgel systems''. Access to
computational resources at the Center for Parallel Computing
at the M.V. Lomonosov Moscow State University
(``Lomonosov'' and ``Chebyshev'' supercomputers) is gratefully
acknowledged.

\footnotesize{

\begin{mcitethebibliography}{52}
\providecommand*{\natexlab}[1]{#1}
\providecommand*{\mciteSetBstSublistMode}[1]{}
\providecommand*{\mciteSetBstMaxWidthForm}[2]{}
\providecommand*{\mciteBstWouldAddEndPuncttrue}
  {\def\EndOfBibitem{\unskip.}}
\providecommand*{\mciteBstWouldAddEndPunctfalse}
  {\let\EndOfBibitem\relax}
\providecommand*{\mciteSetBstMidEndSepPunct}[3]{}
\providecommand*{\mciteSetBstSublistLabelBeginEnd}[3]{}
\providecommand*{\EndOfBibitem}{}
\mciteSetBstSublistMode{f}
\mciteSetBstMaxWidthForm{subitem}
{(\emph{\alph{mcitesubitemcount}})}
\mciteSetBstSublistLabelBeginEnd{\mcitemaxwidthsubitemform\space}
{\relax}{\relax}

\bibitem[Israelachvili(2011)]{Israelachvili2011}
J.~N. Israelachvili, \emph{Intermolecular and Surface Forces}, Academic Press,
  San Diego, 3rd edn, 2011, pp. 291--340\relax
\mciteBstWouldAddEndPuncttrue
\mciteSetBstMidEndSepPunct{\mcitedefaultmidpunct}
{\mcitedefaultendpunct}{\mcitedefaultseppunct}\relax
\EndOfBibitem
\bibitem[Simons and Toomre(2000)]{Simons2000:lipid_rafts}
K.~Simons and D.~Toomre, \emph{Nat. Rev. Mol. Cell Biol.}, 2000, \textbf{1},
  31--39\relax
\mciteBstWouldAddEndPuncttrue
\mciteSetBstMidEndSepPunct{\mcitedefaultmidpunct}
{\mcitedefaultendpunct}{\mcitedefaultseppunct}\relax
\EndOfBibitem
\bibitem[Glotzer and Solomon(2007)]{Glotzer2007}
S.~C. Glotzer and M.~J. Solomon, \emph{Nat Mater}, 2007, \textbf{6},
  557--562\relax
\mciteBstWouldAddEndPuncttrue
\mciteSetBstMidEndSepPunct{\mcitedefaultmidpunct}
{\mcitedefaultendpunct}{\mcitedefaultseppunct}\relax
\EndOfBibitem
\bibitem[Meyer \emph{et~al.}(2005)Meyer, Lin, Hassenkam, Oroudjev, and
  Israelachvili]{Meyer2005:exp_inhomo_force}
E.~E. Meyer, Q.~Lin, T.~Hassenkam, E.~Oroudjev and J.~N. Israelachvili,
  \emph{Proc. Natl. Acad. Sci. U. S. A.}, 2005, \textbf{102}, 6839--6842\relax
\mciteBstWouldAddEndPuncttrue
\mciteSetBstMidEndSepPunct{\mcitedefaultmidpunct}
{\mcitedefaultendpunct}{\mcitedefaultseppunct}\relax
\EndOfBibitem
\bibitem[Popa \emph{et~al.}(2010)Popa, Papastavrou, and
  Borkovec]{borkovec:dendrimers_adsoption}
I.~Popa, G.~Papastavrou and M.~Borkovec, \emph{Phys. Chem. Chem. Phys.}, 2010,
  \textbf{12}, 4863--4871\relax
\mciteBstWouldAddEndPuncttrue
\mciteSetBstMidEndSepPunct{\mcitedefaultmidpunct}
{\mcitedefaultendpunct}{\mcitedefaultseppunct}\relax
\EndOfBibitem
\bibitem[Chen \emph{et~al.}(2006)Chen, Tan, Huang, Ng, Ford, and
  Tong]{LRattraction:Chen2006}
W.~Chen, S.~Tan, Z.~Huang, T.-K. Ng, W.~T. Ford and P.~Tong, \emph{Phys. Rev.
  E}, 2006, \textbf{74}, 021406\relax
\mciteBstWouldAddEndPuncttrue
\mciteSetBstMidEndSepPunct{\mcitedefaultmidpunct}
{\mcitedefaultendpunct}{\mcitedefaultseppunct}\relax
\EndOfBibitem
\bibitem[Walther and Muller(2008)]{Walther2008:Janus}
A.~Walther and A.~H.~E. Muller, \emph{Soft Matter}, 2008, \textbf{4},
  663--668\relax
\mciteBstWouldAddEndPuncttrue
\mciteSetBstMidEndSepPunct{\mcitedefaultmidpunct}
{\mcitedefaultendpunct}{\mcitedefaultseppunct}\relax
\EndOfBibitem
\bibitem[Du and O{'}Reilly(2011)]{Du2011:anisotrop}
J.~Du and R.~K. O{'}Reilly, \emph{Chem. Soc. Rev.}, 2011, \textbf{40},
  2402--2416\relax
\mciteBstWouldAddEndPuncttrue
\mciteSetBstMidEndSepPunct{\mcitedefaultmidpunct}
{\mcitedefaultendpunct}{\mcitedefaultseppunct}\relax
\EndOfBibitem
\bibitem[Pawar and Kretzschmar(2010)]{Pawar2010}
A.~B. Pawar and I.~Kretzschmar, \emph{Macromolecular Rapid Communications},
  2010, \textbf{31}, 150--168\relax
\mciteBstWouldAddEndPuncttrue
\mciteSetBstMidEndSepPunct{\mcitedefaultmidpunct}
{\mcitedefaultendpunct}{\mcitedefaultseppunct}\relax
\EndOfBibitem
\bibitem[Vreeker \emph{et~al.}(1992)Vreeker, Kuin, Boer, Hoekstra, and
  Agterof]{Vreeker1992:stability_hetero}
R.~Vreeker, A.~J. Kuin, D.~C.~D. Boer, L.~L. Hoekstra and W.~G.~M. Agterof,
  \emph{Journal of Colloid and Interface Science}, 1992, \textbf{154}, 138 --
  145\relax
\mciteBstWouldAddEndPuncttrue
\mciteSetBstMidEndSepPunct{\mcitedefaultmidpunct}
{\mcitedefaultendpunct}{\mcitedefaultseppunct}\relax
\EndOfBibitem
\bibitem[Perkin \emph{et~al.}(2006)Perkin, Kampf, and
  Klein]{LRattraction:Perkin2006}
S.~Perkin, N.~Kampf and J.~Klein, \emph{Phys. Rev. Lett.}, 2006, \textbf{96},
  038301\relax
\mciteBstWouldAddEndPuncttrue
\mciteSetBstMidEndSepPunct{\mcitedefaultmidpunct}
{\mcitedefaultendpunct}{\mcitedefaultseppunct}\relax
\EndOfBibitem
\bibitem[Muller and Derjaguin({1983})]{muller.vm:1983}
V.~M. Muller and B.~V. Derjaguin, \emph{Colloids Surfaces}, {1983},
  \textbf{{6}}, {205--220}\relax
\mciteBstWouldAddEndPuncttrue
\mciteSetBstMidEndSepPunct{\mcitedefaultmidpunct}
{\mcitedefaultendpunct}{\mcitedefaultseppunct}\relax
\EndOfBibitem
\bibitem[Miklavic \emph{et~al.}({1994})Miklavic, Chan, White, and
  Healy]{Miklavic:hetero_94}
S.~J. Miklavic, D.~Y.~C. Chan, L.~R. White and T.~W. Healy, \emph{J. Phys.
  Chem.}, {1994}, \textbf{{98}}, {9022--9032}\relax
\mciteBstWouldAddEndPuncttrue
\mciteSetBstMidEndSepPunct{\mcitedefaultmidpunct}
{\mcitedefaultendpunct}{\mcitedefaultseppunct}\relax
\EndOfBibitem
\bibitem[Miklavcic(1995)]{Miklavcic_preturb}
S.~J. Miklavcic, \emph{J. Chem. Phys.}, 1995, \textbf{103}, 4794--4806\relax
\mciteBstWouldAddEndPuncttrue
\mciteSetBstMidEndSepPunct{\mcitedefaultmidpunct}
{\mcitedefaultendpunct}{\mcitedefaultseppunct}\relax
\EndOfBibitem
\bibitem[Ben-Yaakov \emph{et~al.}(2013)Ben-Yaakov, Andelman, and
  Diamant]{Ben-Yaakov2013:hetero}
D.~Ben-Yaakov, D.~Andelman and H.~Diamant, \emph{Physical Review E}, 2013,
  \textbf{87}, 022402\relax
\mciteBstWouldAddEndPuncttrue
\mciteSetBstMidEndSepPunct{\mcitedefaultmidpunct}
{\mcitedefaultendpunct}{\mcitedefaultseppunct}\relax
\EndOfBibitem
\bibitem[Brewster \emph{et~al.}(2008)Brewster, Pincus, and
  Safran]{brewster2008:mobile_heterogeneous}
R.~Brewster, P.~A. Pincus and S.~A. Safran, \emph{Phys. Rev. Lett.}, 2008,
  \textbf{101}, 128101\relax
\mciteBstWouldAddEndPuncttrue
\mciteSetBstMidEndSepPunct{\mcitedefaultmidpunct}
{\mcitedefaultendpunct}{\mcitedefaultseppunct}\relax
\EndOfBibitem
\bibitem[Jho \emph{et~al.}(2011)Jho, Brewster, Safran, and
  Pincus]{pincus2011:heterogeneous_membranes}
Y.~S. Jho, R.~Brewster, S.~A. Safran and P.~A. Pincus, \emph{Langmuir}, 2011,
  \textbf{27}, 4439--4446\relax
\mciteBstWouldAddEndPuncttrue
\mciteSetBstMidEndSepPunct{\mcitedefaultmidpunct}
{\mcitedefaultendpunct}{\mcitedefaultseppunct}\relax
\EndOfBibitem
\bibitem[Boon and van Roij(2011)]{Chg_regulation:van_Roij}
N.~Boon and R.~van Roij, \emph{J. Chem. Phys.}, 2011, \textbf{134},
  054706\relax
\mciteBstWouldAddEndPuncttrue
\mciteSetBstMidEndSepPunct{\mcitedefaultmidpunct}
{\mcitedefaultendpunct}{\mcitedefaultseppunct}\relax
\EndOfBibitem
\bibitem[Lindemann and Winterhalter(2006)]{lindemann.m:2006}
M.~Lindemann and M.~Winterhalter, \emph{IEE Proc. Sys. Biol.}, 2006,
  \textbf{153}, 107\relax
\mciteBstWouldAddEndPuncttrue
\mciteSetBstMidEndSepPunct{\mcitedefaultmidpunct}
{\mcitedefaultendpunct}{\mcitedefaultseppunct}\relax
\EndOfBibitem
\bibitem[Donath \emph{et~al.}(1998)Donath, Sukhorukov, Caruso, Davis, and
  M\"ohwald]{donath.e:1998}
E.~Donath, G.~B. Sukhorukov, F.~Caruso, S.~A. Davis and H.~M\"ohwald,
  \emph{Angew. Chem.-Int. Edit.}, 1998, \textbf{37}, 2202--2205\relax
\mciteBstWouldAddEndPuncttrue
\mciteSetBstMidEndSepPunct{\mcitedefaultmidpunct}
{\mcitedefaultendpunct}{\mcitedefaultseppunct}\relax
\EndOfBibitem
\bibitem[Vinogradova \emph{et~al.}(2006)Vinogradova, Lebedeva, and
  Kim]{vinogradova.oi:2006}
O.~I. Vinogradova, O.~V. Lebedeva and B.~S. Kim, \emph{Ann. Rev. Mater. Res.},
  2006, \textbf{36}, 143\relax
\mciteBstWouldAddEndPuncttrue
\mciteSetBstMidEndSepPunct{\mcitedefaultmidpunct}
{\mcitedefaultendpunct}{\mcitedefaultseppunct}\relax
\EndOfBibitem
\bibitem[Kim \emph{et~al.}(2007)Kim, Lobaskin, Tsekov, and
  Vinogradova]{kim.bs:2007}
B.~S. Kim, V.~Lobaskin, R.~Tsekov and O.~I. Vinogradova, \emph{J. Chem. Phys.},
  2007, \textbf{126}, 244901\relax
\mciteBstWouldAddEndPuncttrue
\mciteSetBstMidEndSepPunct{\mcitedefaultmidpunct}
{\mcitedefaultendpunct}{\mcitedefaultseppunct}\relax
\EndOfBibitem
\bibitem[Vinogradova \emph{et~al.}(2004)Vinogradova, Andrienko, Lulevich,
  Nordschild, and Sukhorukov]{vinogradova.oi:2004}
O.~I. Vinogradova, D.~Andrienko, V.~V. Lulevich, S.~Nordschild and G.~B.
  Sukhorukov, \emph{Macromolecules}, 2004, \textbf{37}, 1113\relax
\mciteBstWouldAddEndPuncttrue
\mciteSetBstMidEndSepPunct{\mcitedefaultmidpunct}
{\mcitedefaultendpunct}{\mcitedefaultseppunct}\relax
\EndOfBibitem
\bibitem[Odijk and Slok(2003)]{odijk.t:2003}
T.~Odijk and F.~Slok, \emph{J. Phys. Chem. B}, 2003, \textbf{107},
  8074--8077\relax
\mciteBstWouldAddEndPuncttrue
\mciteSetBstMidEndSepPunct{\mcitedefaultmidpunct}
{\mcitedefaultendpunct}{\mcitedefaultseppunct}\relax
\EndOfBibitem
\bibitem[Alberts \emph{et~al.}(1983)Alberts, Bray, Lewis, Raff, Roberts, and
  Watson]{alberts.b:1983}
B.~Alberts, D.~Bray, J.~Lewis, M.~Raff, K.~Roberts and J.~D. Watson,
  \emph{Molecular Biology of the Cell}, Garland Publishing, NY \& London,
  1983\relax
\mciteBstWouldAddEndPuncttrue
\mciteSetBstMidEndSepPunct{\mcitedefaultmidpunct}
{\mcitedefaultendpunct}{\mcitedefaultseppunct}\relax
\EndOfBibitem
\bibitem[Sen \emph{et~al.}(1988)Sen, Hellman, and Nikaido]{sen.k:1988}
K.~Sen, J.~Hellman and H.~Nikaido, \emph{J. Biol. Chem.}, 1988, \textbf{263},
  1182\relax
\mciteBstWouldAddEndPuncttrue
\mciteSetBstMidEndSepPunct{\mcitedefaultmidpunct}
{\mcitedefaultendpunct}{\mcitedefaultseppunct}\relax
\EndOfBibitem
\bibitem[Stock \emph{et~al.}(1977)Stock, Rauch, and Roseman]{stock.jb:1977}
J.~B. Stock, B.~Rauch and S.~Roseman, \emph{J. Biol. Chem.}, 1977,
  \textbf{252}, 7850\relax
\mciteBstWouldAddEndPuncttrue
\mciteSetBstMidEndSepPunct{\mcitedefaultmidpunct}
{\mcitedefaultendpunct}{\mcitedefaultseppunct}\relax
\EndOfBibitem
\bibitem[Sukharev \emph{et~al.}(2001)Sukharev, Betanzos, Chiang, and
  Guy]{sukharev.s:2001}
S.~Sukharev, M.~Betanzos, C.-S. Chiang and H.~R. Guy, \emph{Nature}, 2001,
  \textbf{409}, 720\relax
\mciteBstWouldAddEndPuncttrue
\mciteSetBstMidEndSepPunct{\mcitedefaultmidpunct}
{\mcitedefaultendpunct}{\mcitedefaultseppunct}\relax
\EndOfBibitem
\bibitem[Zhou and Stell(1988)]{zhou.y:1988}
Y.~Zhou and G.~Stell, \emph{J. Chem. Phys.}, 1988, \textbf{89}, 7010\relax
\mciteBstWouldAddEndPuncttrue
\mciteSetBstMidEndSepPunct{\mcitedefaultmidpunct}
{\mcitedefaultendpunct}{\mcitedefaultseppunct}\relax
\EndOfBibitem
\bibitem[Deserno and von Gr\"unberg(2002)]{deserno.m:2002}
M.~Deserno and H.~H. von Gr\"unberg, \emph{Phys. Rev. E}, 2002, \textbf{66},
  011401\relax
\mciteBstWouldAddEndPuncttrue
\mciteSetBstMidEndSepPunct{\mcitedefaultmidpunct}
{\mcitedefaultendpunct}{\mcitedefaultseppunct}\relax
\EndOfBibitem
\bibitem[Tsekov and Vinogradova(2007)]{tsekov.r:2006}
R.~Tsekov and O.~I. Vinogradova, \emph{J. Chem. Phys.}, 2007, \textbf{126},
  094901\relax
\mciteBstWouldAddEndPuncttrue
\mciteSetBstMidEndSepPunct{\mcitedefaultmidpunct}
{\mcitedefaultendpunct}{\mcitedefaultseppunct}\relax
\EndOfBibitem
\bibitem[Stukan \emph{et~al.}(2006)Stukan, Lobaskin, Holm, and
  Vinogradova]{stukan.mr:2006}
M.~R. Stukan, V.~Lobaskin, C.~Holm and O.~I. Vinogradova, \emph{Phys. Rev. E},
  2006, \textbf{73}, 021801\relax
\mciteBstWouldAddEndPuncttrue
\mciteSetBstMidEndSepPunct{\mcitedefaultmidpunct}
{\mcitedefaultendpunct}{\mcitedefaultseppunct}\relax
\EndOfBibitem
\bibitem[Tsekov \emph{et~al.}(2008)Tsekov, Stukan, and
  Vinogradova]{tsekov.r:2008}
R.~Tsekov, M.~R. Stukan and O.~I. Vinogradova, \emph{J. Chem. Phys.}, 2008,
  \textbf{129}, 244707\relax
\mciteBstWouldAddEndPuncttrue
\mciteSetBstMidEndSepPunct{\mcitedefaultmidpunct}
{\mcitedefaultendpunct}{\mcitedefaultseppunct}\relax
\EndOfBibitem
\bibitem[Siber \emph{et~al.}(2012)Siber, Bozic, and Podgornik]{siber.a:2012}
A.~Siber, A.~L. Bozic and R.~Podgornik, \emph{Phys. Chem. Chem. Phys.}, 2012,
  \textbf{14}, 3746--3765\relax
\mciteBstWouldAddEndPuncttrue
\mciteSetBstMidEndSepPunct{\mcitedefaultmidpunct}
{\mcitedefaultendpunct}{\mcitedefaultseppunct}\relax
\EndOfBibitem
\bibitem[Ninham and Parsegian(1971)]{nimham.bw:1971}
B.~W. Ninham and V.~A. Parsegian, \emph{J. Theor. Biol.}, 1971, \textbf{31},
  405--428\relax
\mciteBstWouldAddEndPuncttrue
\mciteSetBstMidEndSepPunct{\mcitedefaultmidpunct}
{\mcitedefaultendpunct}{\mcitedefaultseppunct}\relax
\EndOfBibitem
\bibitem[Vinogradova \emph{et~al.}(2012)Vinogradova, Bocquet, Bogdanov, Tsekov,
  and Lobaskin]{NLPB_sim2012}
O.~I. Vinogradova, L.~Bocquet, A.~N. Bogdanov, R.~Tsekov and V.~Lobaskin,
  \emph{J.Chem.Phys.}, 2012, \textbf{136}, 034902\relax
\mciteBstWouldAddEndPuncttrue
\mciteSetBstMidEndSepPunct{\mcitedefaultmidpunct}
{\mcitedefaultendpunct}{\mcitedefaultseppunct}\relax
\EndOfBibitem
\bibitem[Lobaskin \emph{et~al.}(2012)Lobaskin, Bogdanov, and
  Vinogradova]{lobaskin.v:2012}
V.~A. Lobaskin, A.~N. Bogdanov and O.~I. Vinogradova, \emph{Soft Matter}, 2012,
  \textbf{8}, 9428--9435\relax
\mciteBstWouldAddEndPuncttrue
\mciteSetBstMidEndSepPunct{\mcitedefaultmidpunct}
{\mcitedefaultendpunct}{\mcitedefaultseppunct}\relax
\EndOfBibitem
\bibitem[Andelman(2006)]{Nato_andelman}
D.~Andelman, in \emph{Soft Condensed Matter Physics in Molecular and Cell
  Biology}, ed. W.~Poon and D.~Andelman, Taylor \& Francis, New York, 2006,
  ch.~6\relax
\mciteBstWouldAddEndPuncttrue
\mciteSetBstMidEndSepPunct{\mcitedefaultmidpunct}
{\mcitedefaultendpunct}{\mcitedefaultseppunct}\relax
\EndOfBibitem
\bibitem[Ohshima(2010)]{Ohshima2010:plane_out}
H.~Ohshima, \emph{J. Colloid Interface Sci.}, 2010, \textbf{350}, 249 --
  252\relax
\mciteBstWouldAddEndPuncttrue
\mciteSetBstMidEndSepPunct{\mcitedefaultmidpunct}
{\mcitedefaultendpunct}{\mcitedefaultseppunct}\relax
\EndOfBibitem
\bibitem[Deserno and von Gr\"unberg(2002)]{Deserno2002:LPB_unified_approach}
M.~Deserno and H.-H. von Gr\"unberg, \emph{Phys. Rev. E}, 2002, \textbf{66},
  011401\relax
\mciteBstWouldAddEndPuncttrue
\mciteSetBstMidEndSepPunct{\mcitedefaultmidpunct}
{\mcitedefaultendpunct}{\mcitedefaultseppunct}\relax
\EndOfBibitem
\bibitem[Ben-Yaakov and Andelman(2010)]{Andelman:2010_PB_revisit}
D.~Ben-Yaakov and D.~Andelman, \emph{Phys. Stat. Mech. Appl.}, 2010,
  \textbf{389}, 2956 -- 2961\relax
\mciteBstWouldAddEndPuncttrue
\mciteSetBstMidEndSepPunct{\mcitedefaultmidpunct}
{\mcitedefaultendpunct}{\mcitedefaultseppunct}\relax
\EndOfBibitem
\bibitem[Andelman(1995)]{Andelman_biophysics}
D.~Andelman, \emph{Handbook of Biological Physics}, North-Holland, 1995, pp.
  603 -- 642\relax
\mciteBstWouldAddEndPuncttrue
\mciteSetBstMidEndSepPunct{\mcitedefaultmidpunct}
{\mcitedefaultendpunct}{\mcitedefaultseppunct}\relax
\EndOfBibitem
\bibitem[Russel and Schowalter(1989)]{Russel1989}
S.~D.~A. Russel, W.~B. and W.~R. Schowalter, \emph{"Colloidal Dispersions"},
  Cambridge Univ. Press, Cambridge, 1989\relax
\mciteBstWouldAddEndPuncttrue
\mciteSetBstMidEndSepPunct{\mcitedefaultmidpunct}
{\mcitedefaultendpunct}{\mcitedefaultseppunct}\relax
\EndOfBibitem
\bibitem[Bhattacharjee and Elimelech(1997)]{SEI97}
S.~Bhattacharjee and M.~Elimelech, \emph{J. Chem. Phys.}, 1997, \textbf{193},
  273--285\relax
\mciteBstWouldAddEndPuncttrue
\mciteSetBstMidEndSepPunct{\mcitedefaultmidpunct}
{\mcitedefaultendpunct}{\mcitedefaultseppunct}\relax
\EndOfBibitem
\bibitem[Stankovich and Carnie(1996)]{stankovich1996}
J.~Stankovich and S.~L. Carnie, \emph{Langmuir}, 1996, \textbf{12},
  1453--1461\relax
\mciteBstWouldAddEndPuncttrue
\mciteSetBstMidEndSepPunct{\mcitedefaultmidpunct}
{\mcitedefaultendpunct}{\mcitedefaultseppunct}\relax
\EndOfBibitem
\bibitem[Limbach \emph{et~al.}(2006)Limbach, Arnold, Mann, and Holm]{Espresso}
H.~J. Limbach, A.~Arnold, B.~A. Mann and C.~Holm, \emph{Comput. Phys. Commun.},
  2006, \textbf{174}, 704 -- 727\relax
\mciteBstWouldAddEndPuncttrue
\mciteSetBstMidEndSepPunct{\mcitedefaultmidpunct}
{\mcitedefaultendpunct}{\mcitedefaultseppunct}\relax
\EndOfBibitem
\bibitem[Weeks \emph{et~al.}(1971)Weeks, Chandler, and Andersen]{weeks:5237}
J.~D. Weeks, D.~Chandler and H.~C. Andersen, \emph{The Journal of Chemical
  Physics}, 1971, \textbf{54}, 5237--5247\relax
\mciteBstWouldAddEndPuncttrue
\mciteSetBstMidEndSepPunct{\mcitedefaultmidpunct}
{\mcitedefaultendpunct}{\mcitedefaultseppunct}\relax
\EndOfBibitem
\bibitem[Hockney and Eastwood(1989)]{P3M1989}
R.~W. Hockney and J.~W. Eastwood, \emph{{Computer Simulation Using Particles}},
  Taylor \& Francis, 1989\relax
\mciteBstWouldAddEndPuncttrue
\mciteSetBstMidEndSepPunct{\mcitedefaultmidpunct}
{\mcitedefaultendpunct}{\mcitedefaultseppunct}\relax
\EndOfBibitem
\bibitem[Arnold \emph{et~al.}(2002)Arnold, de~Joannis, and Holm]{ELC2002}
A.~Arnold, J.~de~Joannis and C.~Holm, \emph{The Journal of Chemical Physics},
  2002, \textbf{117}, 2496--2502\relax
\mciteBstWouldAddEndPuncttrue
\mciteSetBstMidEndSepPunct{\mcitedefaultmidpunct}
{\mcitedefaultendpunct}{\mcitedefaultseppunct}\relax
\EndOfBibitem
\bibitem[Kuin(1990)]{Kuin1990:superposition}
A.~J. Kuin, \emph{Faraday Discuss. Chem. Soc.}, 1990, \textbf{90},
  235--244\relax
\mciteBstWouldAddEndPuncttrue
\mciteSetBstMidEndSepPunct{\mcitedefaultmidpunct}
{\mcitedefaultendpunct}{\mcitedefaultseppunct}\relax
\EndOfBibitem
\bibitem[Kalasin \emph{et~al.}(2010)Kalasin, Dabkowski, Nüsslein, and
  Santore]{Kalasin2010:bacteri_hetero}
S.~Kalasin, J.~Dabkowski, K.~Nüsslein and M.~M. Santore, \emph{Colloids and
  Surfaces B: Biointerfaces}, 2010, \textbf{76}, 489 -- 495\relax
\mciteBstWouldAddEndPuncttrue
\mciteSetBstMidEndSepPunct{\mcitedefaultmidpunct}
{\mcitedefaultendpunct}{\mcitedefaultseppunct}\relax
\EndOfBibitem
\bibitem[Gingell(1967)]{Gingell1967:surface_potential_cell_interaction}
D.~Gingell, \emph{J. Theor. Biol.}, 1967, \textbf{17}, 451 -- 482\relax
\mciteBstWouldAddEndPuncttrue
\mciteSetBstMidEndSepPunct{\mcitedefaultmidpunct}
{\mcitedefaultendpunct}{\mcitedefaultseppunct}\relax
\EndOfBibitem
\end{mcitethebibliography}
\providecommand*{\mcitethebibliography}{\thebibliography}
\csname @ifundefined\endcsname{endmcitethebibliography}
{\let\endmcitethebibliography\endthebibliography}{}

\bibliographystyle{rsc}
}

\end{document}